\documentclass[preprint,12pt]{elsarticle}
\usepackage{amssymb,bm,graphicx}

\journal{ }

\begin{document}

\begin{frontmatter}

\title{Gamma Ray Spectrum Unfolding Using Derivative Kernels}

\author[UoP]{D. S. Vlachos}
\ead{dvlachos@uop.gr}

\author[UoP]{O. T. Kosmas}
\ead{odykosm@uop.gr}

\address[UoP]{Laboratory of Computer Sciences,\\
Department of Computer Science and Technology,\\
Faculty of Sciences and Technology, University of Peloponnese\\
GR-22 100 Tripolis, Terma Karaiskaki, GREECE}

\begin{abstract}
The unfolding of a gamma ray spectrum experience many difficulties
due to noise in the recorded data, that is based mainly on the change of
photon energy due to scattering mechanisms (either in the detector
or the medium), the accumulation of recorded counts in a fixed
energy interval (the channel width of the detector) and finally the
statistical fluctuation inside the detector. In order to deal with
these problems, a new method is developed which interpolates the
ideal spectrum with the use of special designed derivative kernels.
Preliminary simulation results are presented and show that this
approach is very effective even in spectra with low statistics.
\end{abstract}

\begin{keyword}
Spectrum unfolding \sep Derivative kernels
\PACS 29.30.Kv \sep 29.85.-c \sep 02.60.Jh
\end{keyword}

\end{frontmatter}

\section{Introduction}
\label{sec:intro}
Response function calculation of NaI based scintillators has many applications like process control tasks in manufacturing industry, in oil detection, in safety and alarm systems, in Prompt Gamma Neutron Activation Analysis (PGNAA) and others (see for example \cite{hakimabad_ARI_65_07_918},\cite{tickner_ARI_53_00_507}, \cite{nafee_ARI_xx_08_xx} and references there in). In general, Monte Carlo techniques are used to calculate the interactions of source photons with the detector (\cite{mitra_ARI_63_05_415}, \cite{yalcin_ARI_65_07_1179}, \cite{cengiz_ARI_xx_08_xx}) and thus  the response function. Also both analytical (\cite{abbas_ARI_55_01_245}, \cite{abbas_ARI_64_06_1057}) and statistical (\cite{sabharwal_ARI_xx_08_xx} techniques have been used too.

On the other hand, during the detection of gamma rays, several problems are encountered, ie.
the efficiency vs. resolution of semiconductor or scintillation
detectors used, the geometry, which causes in turn uncertainties in
the solid-angle determination, etc.  Also the form of the spectrum
becomes more complex due to the following properties: (i) the
scintillation detectors have a lower energy resolution compared to
Ge detectors, (ii) the environment and/or shielding play an
important role because of the scattering of high energy x-rays into
the detector, (iii) the first and second escape peaks become
important at high energies and (iv) a significant tail develops
towards the low-energy continuum due to Compton scattering and
escape of bremsstrahlung from the detector. These effects reduce the
detection efficiency in the full-energy peak, and have also other
serious consequences. If the spectrum is complex, with a continuous
$\gamma$-yield (e.g. due to statistical $\gamma$-rays following the
decay of highly excited nuclei), the large superimposing continuous
tails of the high-energy $\gamma$-rays may hamper an accurate
evaluation of the continuous $\gamma$-yield.

To improve these drawbacks several attempts have been made in the past. In the
experiments a combination of different detectors (Ge and BaF,
anti-Compton shields, etc.) has been used. However, these techniques
either reduce the overall efficiency by rejecting a large part of
the detected events (anti-Compton), or hamper a precise
determination of the overall efficiency (addition of coincident
signals from different types of detectors). In the data analysis the
generally used forward method fits the measured spectrum using
appropriate physical models (input information): a master-spectrum
is generated using e.g. statistical model calculations (some model
parameters are to be adjusted later), which is then folded with the
detector response function and the resulting spectrum is compared
with observation. Finally, the model parameters are adjusted, until
an acceptable agreement is found. Problems arise here from peaks in
the experimental spectrum due to contaminants in the target creating
discrete lines, which cannot be simulated easily. The remaining
problem is the choice of the physical model and the appropriate
model parameters. If several physical processes compete, the
generation of the master spectrum can often be ambiguous
\cite{sukosd_NIMPR_355_95_552}. On the other hand, even if the model spectrum is
accurate, the accuracy of the unfolding process is reduced due to
two main reasons: (i) the noise in the measuring spectrum and (ii)
the fact that the measuring spectrum represents the total counts
recorder in a finite energy interval, which is the channel width of
the detector.

The purpose of this work is to present a new method which can improve the unfolding procedure of a given measured spectrum. The method interpolates the
ideal spectrum with the use of special designed derivative kernels.
Preliminary simulation results are presented which show that this
approach is very effective even in spectra with low statistics.

\section{Derivative kernels in unfolding procedure}
\label{sec:main}
\begin{figure}
  \includegraphics[width=\textwidth]{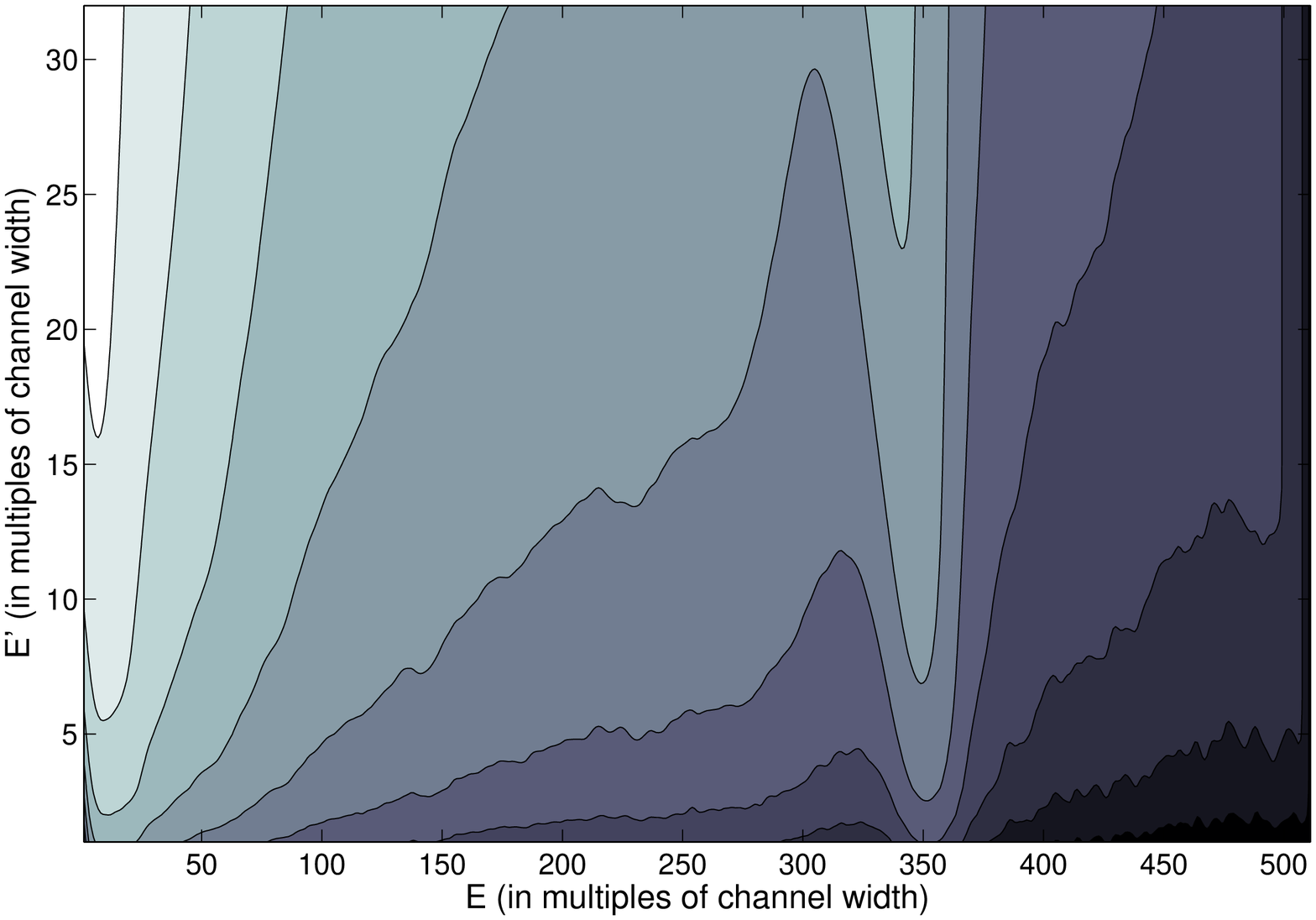}
  \caption{\label{fig_2dim}Two dimensional representation of measured spectrum obtained from \cite{tsabaris_MMS_6_05_35}.}
\end{figure}

\begin{figure}
  \includegraphics[width=\textwidth]{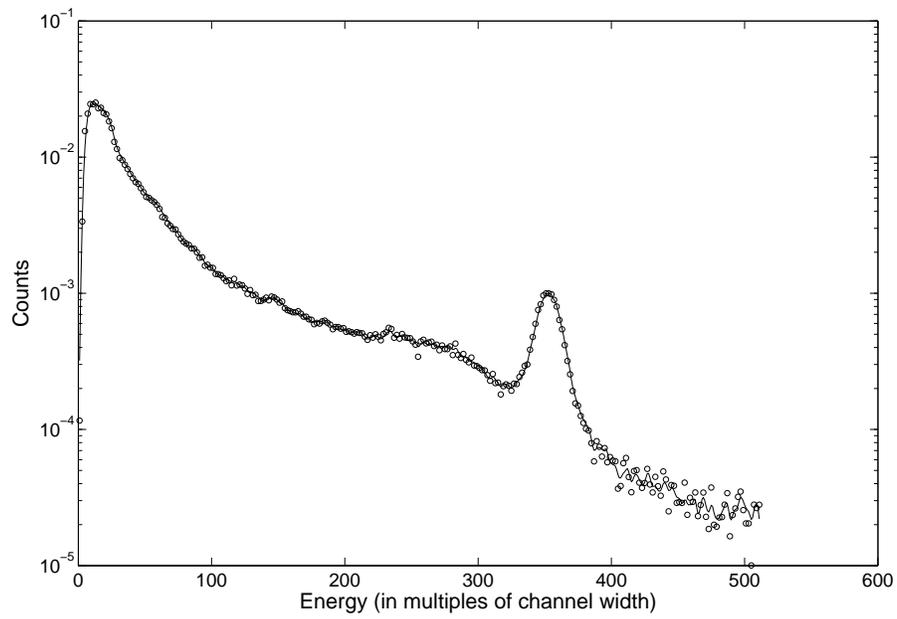}
  \caption{\label{fig_both}Measured ($\circ$) and calculated spectrum (solid line) obtained from \cite{tsabaris_MMS_6_05_35}.}
\end{figure}

Consider the case where a radioactive source emits photons in a
uniform medium and at a given point a photon detector has been
placed. Photons, after their emission and before they reach the
detector, interact with the atoms of the uniform medium and can
change their energy due to Compton scattering or pair production, or
disappear due to the photoelectric effect. The effect of the
interaction of photons with the medium can be formulated as follows.
Let $S(E)$ be the source spectrum and $M(E)$ the measured one. In
vacuum,
\begin{equation}
M(E)=\int _0^{\infty} R(E,V)\cdot S(V)\cdot dV
\end{equation}
where $R(E,V)$ is equal to the number of photons that will be
recorded at energy $E$ when one photon is emitted with energy $V$.
The function $R(E,V)$ is known as the transfer function of the
detector. In the uniform medium, this relation is more complicated.
If a photon with initial energy $U$ is emitted, then there is a
probability $P(V,U)$ that the photon will reach the detector surface
with a final energy $V$. Thus, the measured spectrum now will be
given by:
\begin{equation}
M(E)=\int _0^{\infty} R(E,V)\cdot \left( \int _0^{\infty} P(V,U)
\cdot S(U)\cdot dU \right)\cdot dV
\end{equation}
Changing the order of integration, the function
\begin{equation}
\hat{R}(E,V)= \int _0^{\infty} R(E,U)\cdot P(U,V)\cdot dU
\end{equation}
can be regarded now as the modified transfer function of the
detector, for operation inside the uniform medium \cite{vlachos_JER_82_05_21}. The
measured spectrum can now be expressed as:
\begin{equation}
M(E)=\int _0^{\infty} \hat{R}(E,V)\cdot S(V)\cdot dV
\end{equation}
But instead of the function $M(E)$ the detector integrates this
function in small energy intervals, called channels. Thus, the
detector output $\bar{M}(E)$ is given:
\begin{equation}
\bar{M}(E)=\int _E^{E+\epsilon }M(V)\cdot dV
\end{equation}
where $\epsilon$ is the channel width. Consider now the function
\begin{equation}
m(E,E')=\int _{E}^{E+E'}M(V)\cdot dV
\end{equation}
Since $\bar{M}(E)$ is equal to $m(E,\epsilon )$, the 2-dimensional
function $m(E,E')$ is known on the grid $(n_1\cdot \epsilon ,n_2
\cdot \epsilon ),\; n_1,n_2=0(1)N$. Our purpose now is to find
optimal derivative kernels in order to calculate derivatives of the
function $m(E,E')$. Then, we can calculate $M(E)$:
\begin{equation}
lim _{E'\rightarrow 0} \frac{\partial m(E,E')}{\partial E'}=M(E)
\label{eq_def}
\end{equation}
An important property of $m(E,E')$ which allows for the application
of equation (\ref{eq_def}) is that:
\begin{equation}
m(E,-E')=-m(E-E',E')
\end{equation}
Figure \ref{fig_2dim} shows the measured spectrum from a NaI
detector in an underwater experiment, described in \cite{tsabaris_MMS_6_05_35}.
Both $\bar{M}(E)$ and the calculated from equation (\ref{eq_def})
$M(E)$ are shown in Figure \ref{fig_both}.

The unfolding of the gamma ray spectrum $M(E)$ can now be easily
obtained in the case where radioactive sources emit photons in
discrete energies and the counting rate is low enough to avoid
additive effects in the detector. In this case and based on the
linearity of the folding mechanism, we assume that
\begin{equation}
S(E)=\sum _{n=0}^k a_n \delta (E-E_n)
\end{equation}
and we want to calculate both $a_n$ and $E_n$. Then, it is easily
found that
\begin{equation}
M(E)=\sum _{n=1}^k a_n \hat{R}(E,E_n)
\end{equation}

Finally, consider a continuous function $g:R^2\rightarrow R$ and its
discrete version
\begin{equation}
g_s=\sum _{n_1,n_2 =-\infty}^{\infty}g(x,y)\delta (x-n_1T)\cdot
\delta (y-n_2T)
\end{equation}
where $\delta $ is the Dirac delta function. The knowledge of the
discrete version $g_s$ can lead to the reconstruction of the
continuous function $\bar{g}$ with the aid of a kernel $K$ such that
\begin{equation}
\bar{g}(x,y)=\sum _{n_1,n_2 =-\infty}^{\infty} g_s(n_1,n_2)
K(x-n_1T,y-n_2T)
\end{equation}
The ideal interpolation where $\bar{g}=g$ is achieved if
$K(x,y)=s_T(x)\cdot s_T (y)$, where
\begin{equation}
s_T(x)=\frac{sin(\pi x /T)}{\pi x/T}
\end{equation}
and $T$ is the Nyquist rate. For practical reason, we assume that
\begin{equation}
K(x,y)=d_0(x)\cdot d_0(y)
\end{equation}
and
\begin{equation}
\frac{d^nd_0(x)}{dx^n}=d_n(x)
\end{equation}
Then, the expression for the derivative with respect to x of the
reconstructed function $\bar{g}$ becomes:
\begin{equation}
D_x\{\bar{g}\}(x,y)=\sum _{n_1,n_2
=-\infty}^{\infty}g_s(n_1,n_2)d_1(x-n_1T)d_0(y-n_2T)
\end{equation}
In order to construct an efficient kernel, it is not necessary that
$d_1(x)=d_0'(x)$. Although this seems controversial consider the
following example: it is common to use a sampled Gaussian and its
derivative. However, because the Gaussian is not strictly
bandlimited, sampling introduces artifacts, thus destroying the
derivative relationship between the resulting kernels. So, instead
we choose to simultaneously design a pair of discrete kernels that
optimally preserve the required derivative relationship. If
\begin{equation}
D_0(\omega)=\sum _n d_0(nT)e^{-i\omega n} \;\;,\;\;
D_1(\omega)=\sum _n d_1(nT)e^{-i\omega n} \\
\end{equation}
with $\omega =2\pi /T$ are the discrete Fourier transforms of
$d_0,d_1$, we can demand that
\begin{equation}
i\omega D_0(\omega )=D_1(\omega )
\end{equation}
in the case of one dimensional signals $g$. In the case of two
dimensional signals, we can demand for example that the pair of
kernels preserve the derivative relationship in all directions
\cite{farid_ICCAIP_97}.

\section{Simulation results}
\label{sec:res}
\begin{figure}
  \includegraphics[width=\textwidth]{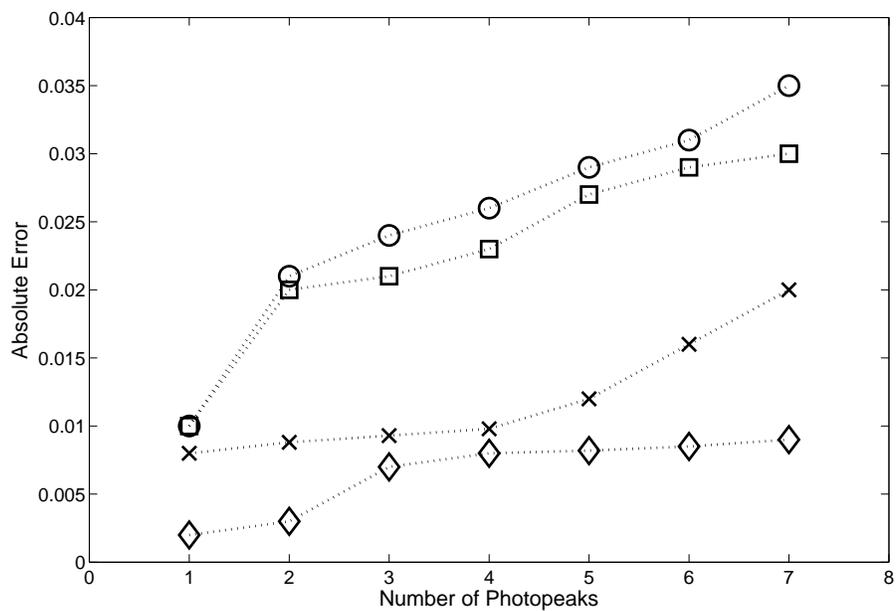}
  \caption{\label{fig_comp1}Unfolding of simulated spectra generated with
  the method in \cite{vlachos_NIMPR_539_05_414}. \emph{DK3} uses the 3-point kernel couple,
  \emph{DK4} the 4-point one and \emph{DK5} the 5-point kernel.
  The \emph{Gaussian} curve is produced using a sampled Gaussian and its derivative. Gaussian
derivative kernel ($\circ$), DK3 ($\Box$), DK4 ($\times$) and DK5 ($\diamond$).}
\end{figure}

\begin{figure}
  \includegraphics[width=\textwidth]{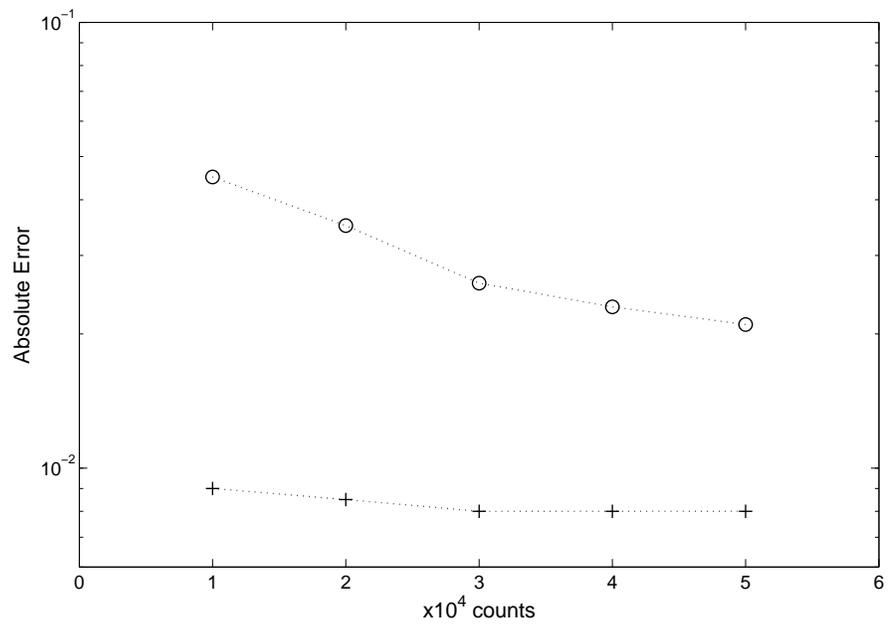}
  \caption{\label{fig_comp2}Dependence of unfolding accuracy on spectrum statistics for the Gaussian ($\circ$) and DK5 ($+$) case. The horizontal axis represents the total number of counts recorded.}
\end{figure}

In order to test the new method, a simulation experiment was
performed. A folded spectrum is produced using the transfer function
for a NaI based measuring system calculated in \cite{vlachos_NIMPR_539_05_414}. The
spectrum is folded again using the method presented in \cite{vlachos_CPC_174_06_391}
in order to simulate an underwater measuring system. Several
simulated spectra were produced, with different number of
photo-peaks and different number of total recorded counts in order
to account for the spectrum statistics. Figure \ref{fig_comp1} shows
the overall error in the unfolded spectrum, using a Gaussian
derivative kernel ($\circ$) and three derivative kernels DK3
($\Box$), DK4 ($\times$) and DK5 ($\diamond$) with 3,4 and 5 points
respectively calculated in \cite{farid_ICCAIP_97}. Furthermore, in Figure
\ref{fig_comp2} the dependence of the error in the unfolded spectrum
on spectrum statistics is shown for the Gaussian ($\circ$) and DK5
($\times$) derivative kernel. It is clear that the new method is
very promising even in cases with low statistics. A special
experimental setup is under construction to test the new method in
real spectrums. Moreover, new derivative kernels are designed in
order to optimize their behavior.

\section{Conclusions}
\label{sec:concl}
Preliminary results on interpolation of a measured spectrum with derivative kernels, show that the unfolding procedure becomes more accurate even in cases of low statistics. The use of derivative kernels facilitate the numerical differentiation which is of high importance in both peak detection and spectrum unfolding.

\section*{Acknowledgments}
This paper is part of the $03ED51$ research project, implemented
within the framework of the "\emph{Reinforcement Programme of Human
Research Manpower}" (PENED) and co-financed by National and
Community Funds ($25\%$ from the Greek Ministry of
Development-General Secretariat of Research and Technology and
$75\%$ from E.U.-European Social Fund).

\bibliographystyle{elsarticle-num}
\bibliography{deriv_ker}

\clearpage

\end{document}